\documentclass[prb,twocolumn, superscriptaddress,showpacs]{revtex4}
\usepackage{amsmath}
\usepackage{version}
\usepackage{graphicx}
\begin{document}

\newcommand{\tab}{\vspace{5mm}\hspace{5mm}}
\setlength{\parindent}{5mm}
\newcommand{\Tc}{T$_{\mathrm C}$}
\newcommand{\sinth}{\mbox{$\sin\theta/\lambda$}}
\newcommand{\inA}{\mbox{\AA$^{-1}$}}
\newcommand{\mub}{\mbox{$\mu_{B}$}}
\newcommand{\mns}{$-$}
\newcommand{\ddd}{3\textit{d} }
\newcommand{\pp}{2\textit{p}}
\newlength{\minusspace}
\settowidth{\minusspace}{$-$}
\newcommand{\msp}{\hspace*{\minusspace}}
\newlength{\zerospace}
\settowidth{\zerospace}{$0$}
\newcommand{\zsp}{\hspace*{\zerospace}}
\newcommand{\lamnsr}{La$_{1.8}$Sr$_{1.2}$Mn$_{2}$O$_7$}
\bibliographystyle{apsrev}


\title{Magnetization Distribution in the layered CMR Manganite La$_{1.2}$Sr$_{1.8}$Mn$_{2}$O$_{7}$
from Polarized Neutron Diffraction}


\author{D. N. Argyriou}
\altaffiliation{Present address: Hahn-Meitner-Institut, Glienicker Stra\ss e 100, D-14 109 Berlin, Germany}
\affiliation{Materials Science Division, Argonne National Laboratory, Argonne, IL, 60439.}
\author{P.J. Brown}
\affiliation{Institut Max Von Laue-Paul Langevin, Boite Postale 156, 38042 Grenoble Cedex 09, France.}
\author{J.Gardner}
\affiliation{NPMR, Chalk River Laboratories, Bld. 459, Stn.18, Chalk River, ON, KOJ 1JO, Canada.}
\author{R. H. Heffner}
\affiliation{MSK764, Los Alamos National Laboratory, Los Alamos, NM, 87545.}



\date{\today}

\begin{abstract}
In the ferromagnetic metallic state of the colossal magnetoresistive (CMR) manganite  La$_{1.2}$Sr$_{1.8}$Mn$_{2}$O$_{7}$, the spin density distribution is essentially in agreement with the standard picture in which the unpaired electrons occupy the three t$_{2g}$ orbitals,
d$_{xy}$, d$_{yz}$ and d$_{xz}$.  However we find a small spin density (\ensuremath{\sim}4\% of the total Mn spin) on the apical O(2) oxygen atom at both 100 K and 220 K and we suggest that this is due to covalency effects.  Surprisingly we find no evidence of spin on the other apical oxygen O(1) suggesting that the Mn e$_{g}$ electron distribution along the \textit{c}-axis is highly anisotropic.
\end{abstract}

\maketitle

\section{Introduction}

In the mixed valent manganite perovskites the electrons in Mn t$_{2g}$ states are treated as localized spins whilst the doped charges occupy e$_{g}$ orbitals and propagate the magnetic coupling either through super or double exchange.\cite{gooden}  However it has recently been realized that Jahn-Teller distortions in the manganites also modulate the relative energies of the different Mn e$_{g}$ orbitals, thus affecting the magnetic coupling in the ordered state.\cite{maez} This effect has been clearly observed in the magnetic properties of the Ruddlesden-Popper  La$_{2-2x}$Sr$_{1+2x}$Mn$_{2}$O$_{7}$ system of naturally layered manganite perovskites that exhibit colossal magnetoresistance at \Tc, much like their three dimensional perovskite $A'_{1-x}A_{x}MnO_{3}$ counterparts.

The magnetic phase diagram of these double layered tetragonal manganites has been examined intensely as a function of electronic doping \textit{x} in recent years  \cite{ling,medarde,kubota,hirota}. An interesting feature of this phase diagram is the rapid change in the magnetic easy axis of the ferromagnetic state over a  relative small range in composition ($\Delta x\sim 0.1$);  the ferromagnetic easy axis varies from easy axis ($\parallel$ \textit{c}) to easy plane ($\parallel$ \textit{a}) as doping varies from $x$=0.4 to 0.3.  These changes have been ascribed both theoretically and experimentally to the energy cross-over between the Mn d$_{3z^{2} -r^{2}}$ orbitals occupied at $x$=0.4 to d$_{x^{2} -y^{2}}$ configuration for $x$=0.3.  Recent calculations by Maezono and Nagaosa \cite{maez3} compute the magnetic phase diagram of these double layered manganites in terms of orbital occupancies, Coulombic repulsion and coherent Jahn-Teller distortions. They predict a range of magnetic structures over this composition range, that are largely in agreement with experimental results. 

This cross-over in the stability of the Mn 3$d$ state between $x$=0.4-0.3 is also evident in the lattice effects that are observed at the coincident electronic and magnetic transition at \Tc.  For example, in the x=0.4 compound a substantial contraction of the \textit{a}-axis (\ensuremath{\sim}0.16\%) and an expansion of the \textit{c}-axis (\ensuremath{\sim}0.06\%) is found \cite{mitchell} suggesting a transfer of charge to Mn d$_{3z^{2} -r^{2}}$ orbitals at \Tc. With decreasing $x$ these effects progressively vary to that of a significant reduction of the  \textit{c}-axis at \Tc \space for compositions up to x=0.3. \cite{medarde}

These experimental and theoretical results highlight the importance of orbital degrees of freedom in the understanding both the structural and long range magnetic properties of these layered CMR manganites. Although unpolarized neutron diffraction measurements allow some inferences to be made about the orbital stability of the Mn \ddd states from the structural deformation of MnO$_{6}$ octahedra they do not measure orbital occupancies directly. In this brief communication we use a single crystal polarized neutron diffraction technique to map out the spin density distribution (unpaired electron density) in the layered CMR manganite  La$_{1.2}$Sr$_{1.8}$Mn$_{2}$O$_{7}$ at 100 K and 220 K and describe it in terms of Mn 3$d$ states, thus providing a more direct measurement that can be compared directly with the above mentioned theoretical predictions.  Our measurements essentially confirm the expected picture of occupied t$_{2g}$ states and provide some evidence for occupied d$_{3z^{2} -r^{2}}$ states for $x$=0.4. In addition we find that there is significant density in one of the apical oxygens O(2) that may be ascribed to covalency effects.

The structure of this manganite consists of a double perovskite layer separated by a simple (La,Sr)O rock salt layer.  The MnO$_{2}$ sheets parallel to the \textit{ab}-plane are made up of four equal Mn-O(3) bonds and are found to be almost flat (Mn-O(3)-Mn\ensuremath{\sim}179 deg.).  Of special note is the asymmetry of the two apical oxygens O(1) and O(2).  The O(1) oxygen is shared by two Mn atoms as it resides between MnO$_{2}$  sheets while the O(2) is bonded to only one Mn and is ionically coordinated with La and Sr atoms in the rocksalt layer.  We find that the Mn-O(2) bond is longer than the Mn-O(1) bond (Mn-O(1)=1.942 {\AA}, Mn-0(2)=1.994 {\AA}  at 300K ).\cite{mitchell}

\section{Experimental}
A single crystal of La$_{1.2}$Sr$_{1.8}$Mn$_{2}$O$_{7}$ was grown using the optical floating zone method. The single crystal used in the 
experiment has dimension of 2x5x7 mm and was cleaved from the boule. The characterization of this crystal included AC susceptibility, which showed a paramagnetic to ferromagnetic transition at 125K. Also, in agreement with previous measurements, the susceptibility parallel to the \textit{a}-axis is higher than that parallel to the \textit{c}-axis, suggesting an easy-plane ferromagnetic arrangement. Single crystal polarized neutron measurements were made on the D3 diffractometer at the Institut Laue-Langevin using a neutron wavelength of 0.843 {\AA}. In this 
configuration the polarization efficiency of the incident neutron beam reflected by the Heusler alloy monochromator is 0.9367(3) when polarized parallel to the applied field direction and  -0.9487(1) parallel to [010] and perpendicular to the \textit{h0l} scattering plane. In our measurements a magnetic field was applied to the sample in a direction perpendicular to the scattering plane, which contained the $\langle110\rangle$ direction. Flipping ratios (FR) from approximately 50 independent reflections with $\sinth<0.6$ \inA\ were measured at 100 and 220 K in magnetic fields of 1 T and 5 T respectively. At 100 K the crystal is in the ordered ferromagnetic state as demonstrated by the temperature dependence of the ferromagnetic (101) reflection at 1T.
 For the 220 K measurement, the thermally disordered moments are aligned by the applied field as shown by the increase in magnetic amplitude of the (101) reflection on raising the field from 1 to 5 T. These two temperatures were chosen as to provide a means of detecting the posibility of a charge redistribution above and below \Tc as suggested by the lattice constants and the recent measurements of Campbell et al.\cite{branton}

\begin{figure}
    \begin{center}
    \resizebox{120mm}{!}{
    \includegraphics*{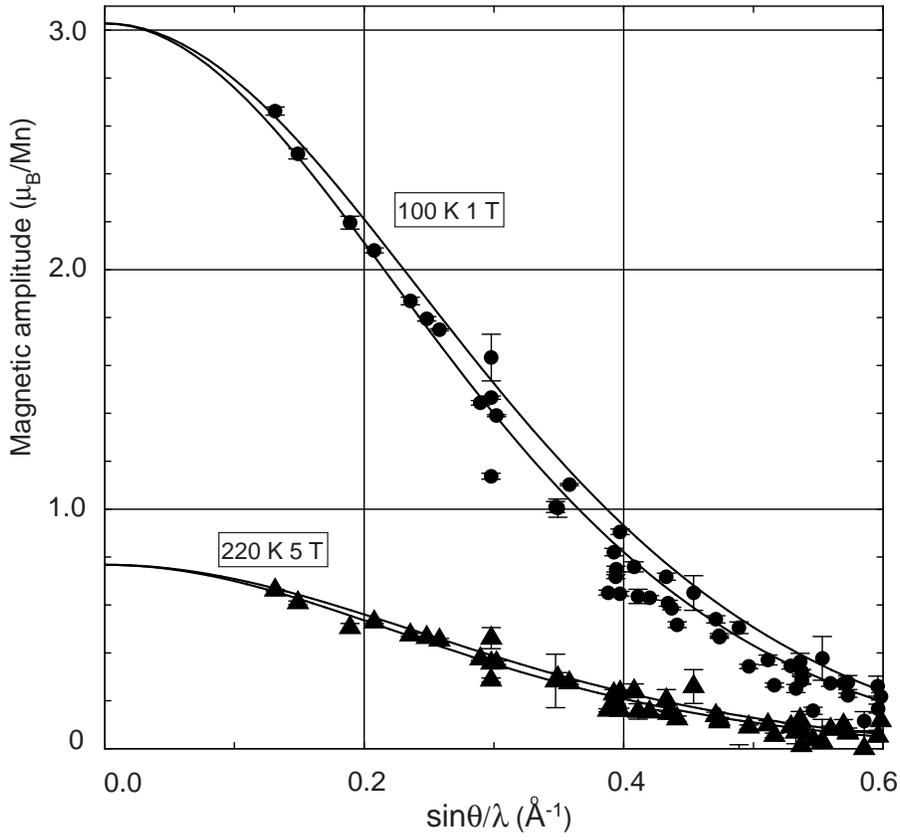}}\\
    \end{center}
    \caption {The magnetic amplitude per Mn atom in \lamnsr\ as a function of \sinth\ at 100 K in 1 T (filled circles) and at 200 K in 5 T (filled triangles). The solid lines are the Mn$^{3+}$(lower) and Mn$^{4+}$(upper) free ion form factors scaled to 0.77 and 3.03 \mub\ for the high and low temperature data respectively.}
    \label{ffact}
\end{figure}

Each FR was obtained by averaging the result of several independent measurements.  The ratio of magnetic to nuclear structure factor ($\gamma$) for each $hkl$ was then calculated from the FR values taking into account geometric effects, lack of complete polarization and extinction.  Nuclear structure factors were computed from the crystallographic parameters reported by Mitchell \textit{et al.}\cite{mitchell}. The contribution of extinction to our measurements was modeled using secondary extinction type I \cite{beckcop}.  This model was quantified from a least squares fit of nuclear intensities measured at 300 K using the time-of-flight single crystal diffractometer at Argonne's intense pulsed neutron source.  The magnetic structure factors F$_{M}$(\textit{hkl}) were obtained by multiplying the $\gamma$ values by the nuclear structure factors.  Standard deviations were calculated from the deviations of measurements of different equivalent reflections from their means. The overall statistical quality of the measured magnetic structure factors is evident from fig.~\ref{ffact}. For the 100K measurement in 1 T, 52 reflections with F$>3\sigma$ were used in our analysis.  Because of the smaller size of the magnetic amplitude at 220K and 5 T, the statistical quality of the data are not as good as for the 100K measurement. In our analysis of these data we used 51 reflections, 42 with  F$>3\sigma$.   

\section{Results}

\begin{figure}
    \begin{center}
    \resizebox{!}{130mm}{
    \includegraphics*{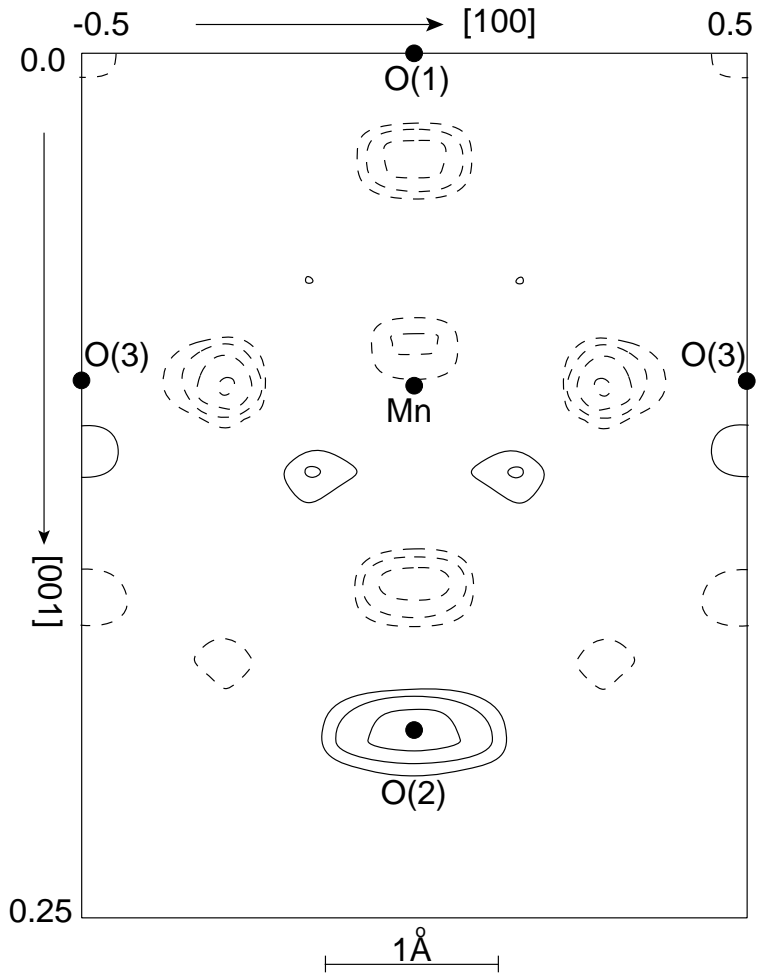}}\\
    \end{center}
    \caption {Maximum entropy reconstruction of the density corresponding to the difference between the observed magnetization distribution and that calculated from the simple spherical model with a Mn$^{3+}$ moment of 3.04 \mub. The section shown is perpendicular to [010] and passes through the origin. The contours are logarithmically spaced with a factor of two between successive levels. The highest contour is at 0.2 \mub \AA$^{-3}$, negative contours are dashed. The positions of atoms in the plane of the section are shown as filled circles.}
    \label{maxent}
\end{figure}

The magnetic amplitude/Mn atom, obtained by dividing the measured magnetic structure factors by the geometric structure factor for the Mn atoms ($4\cos(2\pi l z_{{\rm Mn}}$), is plotted as a function of \sinth\ in fig.~\ref{ffact} for the 100 and 220 K measurements.  The solid lines represent the Q dependence of the spherically symmetric free ion form factors for Mn$^{3+}$ (lower) and Mn$^{4+}$ (upper).  Fitting the measurements at 100 K to a simple Mn$^{3+}$ spherical form factor, we obtain a Mn moment of 3.03(3) \ensuremath{\mu}$_{B}$/Mn.  On the basis of this simple model, a map of the differences between our observation and a spherically symmetric spin distribution on the Mn atom was obtained using the Maximum entropy technique \cite{papo}. The section passing through the Mn and all three O site shows a small negative difference density near the Mn site which is surrounded by larger negative areas displaced along the Mn--O bonds  (fig.~\ref{maxent}). These are just the directions in which the lobes of the e$_g$ functions  point, so the negative difference density probably arises because these functions are missing from the actual density whilst present, together with the t$_{2g}$ functions, in the spherical model. This conclusion is strengthened by the observation of similar regions of positive difference density in $\langle110\rangle$ directions (not visible in fig.~\ref{maxent}), where the maxima in the t$_{2g}$ functions occur. There is, in addition significant density exactly on the apical O(2) oxygen.  The positive density shows that the polarization of O(2) moment is parallel to that of the Mn. We suggest that this feature is related to the bonding between Mn and O, and will be discussed further below.

To arrive at a description of the moment distribution around the Mn ions beyond the simple picture of a spherical density we used two different models, calculating a map of the difference spin density as described above at each turn.  First we used an augmented spherical model in which we allowed both spin and orbital contributions to the \ddd Mn moment, as well as \pp\space spin moment on the oxygen atoms. A second multipolar model was utilized were the moment distribution of the Mn ion was described in terms of the spatial distribution and population of Mn 3$d$ orbitals. 

For the first model the 100 and 220 K data show a significant decrease in the agreement indices ($\chi^{2}$ and R-factor) with the addition of 4 variables (see table~\ref{pars}). The $\chi^{2}$ decreased by $\sim1/3$ for both sets of measurements while the R-factor decrease from 7.8 to 6.6 for the 100K measurement and 12.6 to 11.6 for the 220K.  For the former measurement, the decrease is less statistically significant since the magnetic amplitude at 220K, 5T is much smaller than in the ferromagnetic state. From the 100 K data we compute a spin moment of \ensuremath{\mu}$_{s}$\ensuremath{=}2.78(10) \ensuremath{\mu}$_{B}$/Mn and an orbital moment of  \ensuremath{\mu}$_{\ensuremath{o}}$=0.35(9) \ensuremath{\mu}$_{B}$/Mn.  The total moment (\ensuremath{\mu}$_{s}$+\ensuremath{\mu}$_{\ensuremath{o}}$) is found to be 3.13(5) \ensuremath{\mu}$_{B}$/Mn at 100 K. Interestingly this model gives a moment of 0.18(6) \ensuremath{\mu}$_{B}$  on the apical O(2) atom, while no significant moments are found on the apical O(1)-atom between MnO$_{2}$ sheets or on the in-plane O(3) atom.  For the 220 K data we find similar trends, with the magnitude of the moments scaling with temperature. The improvement in the agreement indices is also reflected with an essentially flat difference density map, although some features still persists between the Mn and O(3) atoms, these being likely due to the aspherical moment distribution of the Mn ions.  

In the second model the asphericity of the moment distribution around the Mn atoms is taken into account.  Here we still allowed for the presence of a spherically distributed moment around the three oxygen sites, but the Mn density was modeled by a sum of spherical harmonic multipole functions allowed by the site symmetry (4\textit{mm}) up to order 4.  The results are given in table~\ref{pars}.  In this analysis the multipolar expansion increases the number of variables to 8, but the reduction of \ensuremath{\chi}$^{2}$ is significant for both 100 K and 220 K measurements. For both measurements,  $\chi^{2}$ decreased by $\sim40\%$, while the R-factor decrease from 6.6  to 5.2 for the 100K measurement and 11.6 to 9.2 for the 220K.  In both cases a statistically significant decrease in the agreement indices. The magnitudes of the multipole terms in the density are linearly related to the occupancies of the symmetrically distinct \ddd orbitals and the values derived from the multipole amplitudes are given in the final block of table~\ref{pars}.  This analysis clearly shows that in the ferromagnetic state 85(10)\% of spin around the Mn is due to electrons in the t$_{2g}$ states ($d_{xy}$ and $d_{zx}+d_{zy}$ from table~\ref{pars}).  The occupancy of the $d_{x^{2} -y^{2} }$ orbitals is not statistically significant while the analysis does suggest a small number of unpaired electrons in the $d_{3z^{2} -r^{2}} $ states for the 100K measurement.  As with the spherical model we find that there is no evidence of a moment at the apical O(1) and in-plane O(3) oxygens. However a significant moment of 0.17(4) was found on the apical O(2) oxygen in accordance with the previous analyses for the 100K measurement.  The same analysis procedure was followed for the measurement at 220 K. Here again we find that 91(11)\% of the moment arises from electrons in the t$_{2g}$ states and a statistically significant moment is still present at the O(2) apical oxygen of 0.06(1) \mub, while there is no evidence of a moment distribution from electrons in the $d_{x^{2} -y^{2} }$ and $d_{3z^{2} -r^{2}}$ states. An essentially flat difference density map was obtained using this model suggesting that the main features in the difference density map shown in fig. ~\ref{maxent} have been accounted for here. 

It is evident from the final values of $\chi^{2}$ that the models used do not give a really satisfactory fit to the data particularly that taken at 100K where the strength of the magnetic scattering allowed high statistical precision to be obtained. It should be emphasized that the high $\chi^{2}$  is due rather to the limitations of the models available with which to fit the data than to problems with the the data themselves. Nonetheless the use of an aspherical spin density as described in the second model dramatically improves the quality of the analysis while providing a physical picture of the magnetization distribution around the Mn ion. 

\section{Discussion}
It is generally accepted that the effect of an octahedral crystal field on Mn ions is to lower the energy of the \ddd states with t$_{2g}$ symmetry with respect to those with  e$_{g}$ symmetry.  Unpaired electrons in the t$_{2g}$ states can thus be considered as localized spins. In the case of Mn$^{3+}$ rich compositions the extra electrons occupy e$_{g}$ states which are split by tetragonal components of the crystal field into non-degenerate  $d_{x^{2} -y^{2} }$ and $d_{3z^{2} -r^{2}}$ orbitals. The detailed properties of the doped manganites depend on  which of the two has the lower energy. Indeed, as has been demonstrated for LaMnO$_{3}$ the redistribution of electrons between these two states is strongly coupled with the magnetic structure. Unfortunately our measurements have proven to lack sensitivity in showing unambiguously the preferential occupation of the $d_{3z^{2} -r^{2}}$ orbital over the $d_{x^{2} -y^{2} }$ below \Tc\space although they do suggest that this picture may possibly be correct.

The observation of a magnetization density on the O(2) deserves some discussion.  Early measurements of magnetization density on carbonates and oxides have reported magnetic moment on oxygen sites \cite{brown00,rackek} which has been attributed to covalency. Indeed Lingard and Marshall\cite{lingard} as early as 1969 predicted theoretically that in the oxide MnCo$_{3}$ there will be an oxygen moment of \ensuremath{\sim}4\% of the size of the Mn$^{2+}$ moment, and \ensuremath{\sim}4\% on the C-atoms. Similar effects have been reported for the single-layer perovskite K$_{2}$CuF$_{4}$. \cite{Ito1,ito2} In the manganite perovskites recent calculations also predict  a moment in the oxygen sites surrounding the Mn-atom \cite{fava, matar, pickett}.  Recently Pierre \textit{et.  al.} \cite{pierre} confirm this experimentally in orthorhombic La$_{0.8}$Sr$_{0.2}$MnO$_{3}$ where they find  a \ensuremath{\sim}0.1\ensuremath{\mu}$_{B}$ moment on the O-atoms using the same polarized neutron beam technique described in this paper.

In the present experiment the observation of a positive magnetization on the O(2) atom and a negative magnetization density along the Mn-O(2) bonds can be understood as due to orbital overlap between Mn e$_{g}$ and oxygen \pp\space orbitals. The transfer of \textit{s}-electrons to the oxygen \pp-shell takes place when the \ddd and \pp\space electrons have similar energies so that hybridization between these two states occurs. This has been confirmed by electronic structure calculations that indicate a large overlap between Mn e$_{g}$ \ddd and Oxygen \pp\space orbitals results in a partially covalent character for the Mn-O bond. Calculations by Freyria-Fava \textit{et al.} \cite{fava} and Pickett and Singh \cite{pickett} predict spin density on the oxygen atoms in perovskite manganites using \textit{ab initio} Hartree-Fock approach or linearized augmented plane-wave (LAPW) methods.   Although both of these theoretical models are in general agreement with our measurements, Pickett and Singh \cite{pickett} predict a smaller moment on the oxygen atom for a similarly doped ferromagnetic manganite perovskite (\ensuremath{\sim} 0.08\mub)  than what we measure here in this layered manganite (0.17\mub). This discrepancy may reflect that the extend of hybridization in these layered manganites is inherently somewhat larger, than the three dimensional perovskites. 

The unusual feature of our measurements is that they suggest that the degree  of covalency in this layered manganites is highly anisotropic  as we find no evidence of an appreciable moment on the apical  oxygen O(1) found between MnO$_{2}$ sheets or along the in-plane oxygen O(3). This suggests that covalency effects are found between only one of the two Mn-O apical bond and that the  d$_{3z^{2} -r^{2}}$ orbital
configuration is anisotropic with higher charge density  in the lobe pointing towards the O(2) apical oxygen.  

\section{Conclusion}
In summary we find that the spin density distribution in the layered manganite La$_{1.2}$Sr$_{1.8}$Mn$_{2}$0$_{7}$ is in
accordance with the standard picture in which the unpaired electrons occupy the three t$_{2g}$ orbitals, d$_{xy}$,
d$_{yz}$ and d$_{xz}$, while there is some evidence to suggest an occupation of the d$_{3z^{2} -r^{2}}$ orbital below \Tc. We find a small spin density (\ensuremath{\sim}4\% of the total Mn spin) on the apical O(2) oxygen atom at both 100 K. Band structure calculations suggest that this additional moment may due to covalency effects. 

\begin{acknowledgments}
We are grateful to E. Leli\`evre-Berna for his help with the D3 diffractometer and Art Schultz for his help on the SCD spectrometer at IPNS. This work was supported by the U.S. Department of Energy, under Contract No. W-31-109-ENG-38 (DNA) and W-7405-ENG-36 (RH). \end{acknowledgments}

\begingroup
\squeezetable
\begin{table}
\caption{ Values of the parameters obtained in least 
squares fits of different models to the magnetic structure factors 
measured for \lamnsr. \label{pars}}\vspace{1ex}
\begin{tabular}{llllllll}
\hline
Data Set&100 K at 1 T&\hspace{20mm}&220 K at 5 T\\
No. of Observations&\msp 52 &&\msp 51\\
\hline
\multicolumn{4}{l}{Spherical Model: 1 variable}\\
${\chi^2}$  \footnotemark[1]&\msp 65&&\msp 8.4\\
$R(\%)$\footnotemark[2]&\msp7.8&&\msp12.6\\
Mn $\langle j0\rangle$&\msp 3.03(3)\footnotemark[3]&&\msp 0.766(7))\\
\hline
\multicolumn{3}{l}{Augmented Spherical Model : 5 variables}\\
$\chi^2$&\msp 41&&\msp 4.9\\
$R(\%)$&\msp 6.6&&\msp 11.6\\
Mn $\langle j0\rangle$&\msp 3.13(5)&&\msp 0.786(8))\\
Mn $\langle j2\rangle$&\mns 0.35(9)&&\mns 0.10(2)\\
O(1)&\mns 0.09(7)&&\msp 0.035(14)\\
O(2)&\msp 0.18(6)&&\msp 0.060(14)\\ 
O(3)&\msp 0.00(3)&&\msp 0.004(5)\\
\hline
\multicolumn{3}{l}{Multipole Model: 8 variables}\\
$\chi^2$&\msp 24&&\msp 3.0\\
$R(\%)$&\msp 5.2&&\msp 9.2\\
Mn $Y(00)$&\msp 3.01(4)&&\msp 0.775(7)\\
Mn $Y(20)$&\msp 0.02(5)&&\msp 0.00(2)\\
Mn $Y(44+)$&\mns 0.5(3)&&\mns 0.15(8)\\
Mn $Y(40)$&\mns 0.5(1)&&\mns 0.18(4)\\
Mn $\langle j2\rangle$&\mns 0.2(1)&&\mns 0.03(3)\\
O(1)&\mns 0.08(5)&&\mns 0.026(11)\\
O(2)&\msp 0.17(4)&&\msp 0.060(11)\\ 
O(3)&\msp 0.01(2)&&\mns 0.002(4)\\
\hline
\multicolumn{3}{l}{Orbital occupancies from multipole model}\\
Mn $d_{x^2-y^2}$&\msp\zsp 5(7)\%&&\msp\zsp 2(8)\%\\
Mn $d_{3z^2-r^2}$&\msp 10(7)\%&&\msp\zsp 7(8)\%\\
Mn$d_{xy}$&\msp 31(7)\%&&\msp 34(8)\%\\
Mn $d_{zx}+d_{zy}$&\msp 54(7)\%&&\msp 58(8)\%\\
Mn g-factor\footnotemark[4]&\msp 1.89(7)&&\msp 1.93(7)\\
\hline
\\
\end{tabular}
\footnotetext[1]{Here $\chi^2 =\Sigma[(F_{obs}-F_{calc})/\sigma(F_{obs})]^2/ (N_{obs}-N_{var})$. The values for the data taken at 220 K are smaller than this for the 100 K data because the magnitude and hence the statistical significance of the structure factors measured at the higher temperature is less good.} 
\footnotetext[2]{$R =100[\Sigma(F_{obs}-F_{calc})/\Sigma(F_{obs})]$.}
\footnotetext[3]{The values of all the variables are given in Bohr magnetons per atom.}
\footnotetext[4]{In the dipole approximation $g = 2(\langle j0\rangle + \langle j2\rangle) /\langle j0\rangle$.}
\end{table}
\endgroup



\end{document}